\documentclass[12pt]{article}
\usepackage{amsmath}
\usepackage{amsthm}
\usepackage{amssymb}
\usepackage{mathrsfs}
\begin{document}
\def\Tm{T^m}
\def\tm{t_m}
\def\Tdm{T^m_d}
\def\tdm{t_m^d}
\def\ot{\otimes}
\def\br{\mathbb{R}}
\def\al{\alpha}
\def\bt{\beta}
\def\th{\theta}
\def\ga{\gamma}
\def\vth{\vartheta}
\def\de{\delta}
\def\lm{\lambda}
\def\b{\beta}
 \def\tp{{\rm tg}({\phi\over 2})}
\def\k{\kappa}
 \def\pp{{\pi\over 2}}
\def\om{\omega}
\def\si{\sigma}
\def\w{\wedge}
\def\od{\sqrt{2}}
\def\Tn{T^n}
\def\tn{t_n}
\def\Tdn{T^n_d}
\def\tdn{t_n^d}
\def\e{\varepsilon}
\def\ti{\tilde}
\def\js{{1\over 4}}
 \def\D{{\cal D}}
\def\I{{\cal I}}
\def\L{{\cal L}}
\def\ri{{\mathrm{i}}}      
\def\S{{\cal S}}
 \def\H{{\cal H}}
\def\G{{\cal G}}
\def\bz{{\bar z}}
\def\E{{\cal E}}
\def\B{{\cal B}}
\def\M{{\cal M}}
 \def\A{{\cal A}}
\def\K{{\cal K}}
\def\J{{\cal J}}
\def\ub{\Upsilon(b)}
\def\tJ{\ti{\cal J}}
\def\stc{\stackrel{.}{+}}
\def\R{{\cal R}}
\def\d{\partial}
\def\la{\langle}
\def\ra{\rangle}
\def\bc{{\mathbb C}}
\def\st{\stackrel{\w}{,}}
\def\lta{\leftarrow}
\def\rta{\rightarrow}
\def\scu{$SL(2,\bc)/SU(2)$~ $WZW$  }
\def\xpm{\partial_{\pm}}
\def\xp{\partial_+}
 \def\xm{\partial_-}
 \def\ps{\partial_{\sigma}}
 \def\1{{\mbox{\boldmath $1$}}}   
  \def\pt{\partial_{\tau}}
\def\be{\begin{equation}}
\def\ee{\end{equation}}
\def\jp{\frac{1}{ 2}}
\def\noi{\noindent}
\def\nl{\nabla}
\def\sve{\sqrt{\vert\e\vert}}
\def\Ad{{\rm Ad}}
 
\def\tD{\Delta^*}
\def\slc{SL(2,{\bf C})}

\begin{titlepage}
\begin{flushright}
{}~
  
\end{flushright}

\vspace{1cm}
\begin{center}
{\large \bf   Poisson-Lie T-duals of the  bi-Yang-Baxter  models}\\ 
[50pt]{\small
{ \bf Ctirad Klim\v{c}\'{\i}k}
\\
Aix Marseille Universit\'e, CNRS, Centrale Marseille\\ I2M, UMR 7373\\ 13453 Marseille, France}
\end{center}

\vspace{0.5 cm}
\centerline{\bf Abstract}
\vspace{0.5 cm}
\noindent   We prove the conjecture of Sfetsos, Siampos and Thompson that suitable analytic continuations of the Poisson-Lie T-duals of the bi-Yang-Baxter sigma models coincide with the recently introduced generalized $\lambda$-models.   We then generalize this result by showing that the analytic continuation of a generic $\sigma$-model of "universal WZW-type" introduced by Tseytlin in 1993 is nothing but the Poisson-Lie T-dual of a generic 
Poisson-Lie symmetric $\sigma$-model introduced by Klim\v c\' ik and \v Severa in 1995.
\vspace{2cm}


\vspace{1cm}


\end{titlepage}

\noindent {\bf 1. Introduction.}
Two kinds of integrable nonlinear $\sigma$-models, the so-called $\eta$-deformation of the principal chiral model \cite{K02,K09} and the $\lambda$-deformation of the WZW model \cite{S14}, 
have recently attracted much attention because of their relevance in string theory  or in non-commutative geometry \cite{strings}.  The integrability of those models was proven at the level of the Lax pair in \cite{K09,S14} and  at the level of the so called $r/s$ exchange relations  in \cite{DMV14}.  Both the $\eta$-model and the $\lambda$-model turned out to be deformable further to give rise to several  families of multi-parametric integrable $\sigma$-models\footnote{The strong integrability in the $r/s$-sense of the so-called bi-Yang-Baxter model of \cite{K14} was further established in \cite{DLMV16a}.}
 \cite{DMV15,K14,SST15} living on general semi-simple group targets (those families  generalize some of the integrable families of $\sigma$-models on low dimensional group targets obtained previously  in \cite{low}).
 
In three recent papers \cite{V15}, \cite{HT} and \cite{SST15}, there was suggested that  the  $\eta$-deformation of the principal chiral model \cite{K02,K09} and the $\lambda$-deformation of the WZW model should be related 
by the  Poisson-Lie T-duality \cite{KS95,KS96}  followed   by an appropriate analytic continuation of the geometry of the $\lm$-model target. In particular, such suggestion was fully worked out for the 
simplest group target $SU(2)$  in \cite{SST15}  where it was shown that the Poisson-Lie T-dual of the bi-Yang-Baxter model \cite{K14} coincides with the analytically continued generalized  $\lambda$-model \cite{SST15}.
Furthermore, Sfetsos, Siampos and Thompson conjectured  that the same result should hold for the bi-Yang-Baxter model living on a  general simple compact group target. We have partially proved this conjecture in \cite{K15}  in the following sense: we did work with the general simple compact group target but we have switched off one of the two deformation parameters of the bi-Yang-Baxter model. Said in other words, we have established in \cite{K15} for every simple compact group target that the Poisson-Lie T-dual of the Yang-Baxter model \cite{K02,K09} coincides with the analytically continued $\lambda$-deformation of the  WZW model \cite{S14}. The first purpose of the present letter is to switch on also the second parameter and, hence, to prove the conjecture of Sfetsos, Siampos and Thompson in its strongest form.

The second purpose of our work is to reveal a highly nontrivial structural relation between two classes of $\sigma$-models  introduced more than twenty years ago:  the  class of "universal WZW-type conformal $\sigma$-models"  introduced by Tseytlin in \cite{T}; we shall refer to them as T-models; and the class of "Poisson-Lie T-dualizable $\sigma$-models on a compact group target" introduced by Klim\v c\' ik and \v Severa in \cite{KS96}; we shall call them KS-models. Namely, we show that the T-models are nothing but the analytic continuations of the Poisson-Lie T-duals of the KS-models.  

By the way, we find truly remarkable that the T-models and the KS-models were orbiting around for two decades without "knowing about each other". The reason for this is that the authors of \cite{KS96} have worked out the target space geometries of the Poisson-Lie T-duals of the KS-models in   coordinates natural from the point of view of Poisson-geometry but not natural for the comparison with the T-model. The parametrization of the dual target suitable for this comparison was introduced in \cite{K15} and here we use it to establish the announced result. We find also interesting that the KS-models were originally invented as new objects, the reason of existence of which was their   T-dualisability of a new kind, and the authors of \cite{KS96} 
were not aware that those models were closely related to the T-models already existing on the market which had their independent reason of existence.  

Our  technical strategy to realize the   first purpose of this work will be the following one: First we represent the bi-Yang-Baxter $\sigma$-model on the target of the simple compact Lie group $G$ as the so-called $\E$-model of Ref. \cite{KS96,KS97,K15} which will permit us to dualize it  in the sense of the Poisson-Lie T-duality. Then we  work out explicitly the resulting dual $\sigma$-model on the target $G^\bc/G$ and we then  establish that  its suitable analytic continuation  coincides
 with the  generalized $\lambda$-model  of Ref.\cite{SST15}.  We then realize  the second purpose by repeating the same procedure for the most general $\E$-model based on the same Drinfeld double. We finish our note with a short outlook.

 \vskip1pc
 
 \noindent {\bf 2.  $\E$-models.}  Recall that the $\E$-model,  introduced in  \cite{KS96,KS97,K15},  is a first-order dynamical system based on a current algebra  of a
 quadratic\footnote{Recall that the quadratic Lie algebra  $\D$ is by definition equipped with a  non-degenerate ad-invariant symmetric bilinear form  $(.,.)_\D$.} Lie algebra $\D$ (playing the role of the symplectic structure)   and  with a  Hamiltonian $H_\E$ being encoded in a choice of a particular linear self-adjoint involution $\E:\D\to\D$.  More precisely, the phase space of the $\E$-model is an infinite-dimensional symplectic  manifold $P_\D$  with a set of distinguished $\D$-valued  coordinates $j(\sigma)$  ($\sigma$ is a  loop parameter)  the  Poisson brackets of which are given by 
\be \{j^A(\sigma), j^B(\sigma')\}=F^{AB}_{~~~C}j^C(\si)\delta(\si-\si')+D^{AB}\d_\si\delta(\si-\si').\label{dca}\ee
Here $F^{AB}_{~~~C}$ are the structure constants of the Lie algebra $\D$ in some basis $T^A\in\D$ and
\be D^{AB}:=(T^A,T^B)_\D,\quad  (j(\si),T^A)_\D:=j^A(\si).\ee
Recall also that the "linear self-adjoint involution" means that  $\E:\D\to\D$ verifies 
\be (\E u,v)_\D=(u,\E v)_\D, \quad \forall u,v\in\D; \qquad \E^2u=u, \quad \forall u\in\D.\ee 
Finally the Hamiltonian of the $\E$-model is given by
\be H_\E:=\jp\int d\si(j(\si),\E j(\si))_\D.\label{ham}\ee

\vskip1pc

\noindent{\bf  3. $\sigma$-models from  the $\E$-models.}  
If there is  a Lie subalgebra $\tilde\G$  of $\D$  of dimensionality  dim$\tilde\G=\jp$dim$\D$ and such that
  $(u,u)_\D=0, \forall u\in\tilde\G$ then for each $\E$ there exists a non-linear $\sigma$-model on the target $D/\tilde G$,  the first order dynamics of which coincides with  the
$\E$-model $(P_\D,H_\E)$.  Here $\tilde G$ and $D$  stand for  (simply connected) Lie groups corresponding to the Lie algebras $\tilde\G$ and $\D$.  
The second order geometrical action of this $D/\tilde G$ model is given by  \cite{KS97,K15,KS97b}:
 \be S_\E(f)=S_{WZW,\D}(f) -k\int d\xi^+ d\xi^-(P_f(\E) f^{-1}\d_+f,f^{-1}\d_-f)_\D,\label{emo}\ee
where  $f\in D$  parametrizes the right coset $D/\tilde G$  (one can choose several local sections covering the whole base space $D/\tilde G$ if there exists no global section of this fibration). 
Most importantly,  the symbol $P_f(\E)$ appearing in \eqref{emo}  denotes a projection operator from $\D$ into $\D$,
unambiguously  defined by the relations
\be {\rm Im}P_f(\E)=\tilde\G,\qquad {\rm Ker}P_f(\E)=(\1+{\rm Ad}_{f^{-1}}\E{\rm Ad}_{f})\D. \label{nd} \ee 
 For completeness,  the standard level $k$ WZW action $S_{WZW,\D}(f)$ is defined as usual 
 $$ S_{WZW,\D}(f):=$$\be :=\frac{k}{2}\int  d\xi^+d\xi^-(f^{-1} \partial_+ f,f^{-1}\partial_-f)_\D+ \frac{k}{12}\int d^{-1}(dff^{-1},[dff^{-1},dff^{-1}])_\D,\ee
and the  light-cone variables $\xi^\pm$ and the derivatives $\d_\pm$  are 
\be \xi^\pm:=\jp(\tau\pm±\sigma), \qquad \d_\pm:=\d_\tau\pm\d_\sigma.\ee

 \vskip1pc
 
\noindent {\bf  4. The  $\E$-model for the Yang-Baxter $\sigma$-model.}
 The question which is often of interest is in a sense inverse to that answered in {\bf 3.} That is,  given a $\sigma$-model on some target, can we associate to it an $\E$-model
 from which it originates via the formula \eqref{emo}?  For example, let us consider the so-called $\eta$-model (or Yang-Baxter $\sigma$-model) \cite{K02,K09}  which is the $\sigma$-model
 on the target of a simple compact group $G$ with the second-order action
\be S_\eta(g)=2k\eta\int  d\xi^+d\xi^-(g^{-1} \partial_+ g,(1-\eta R)^{-1}g^{-1}\partial_-g).\label{0}\ee
Here $g(\xi^+,\xi^-)\in G$ is a field configuration, $(.,.)$ is the  Killing-Cartan form on the Lie algebra $\G^\bc$ of $G^\bc$ and $R:\G\to\G$ is  the so called Yang-Baxter operator defined, for example, in \cite{K09}. 

It turns out (cf. \cite{K02,K09,K15}) that the model \eqref{0} is the $\E$-model for the choice\footnote{$AN$ is the subgroup of $G^\bc$ featuring in the Iwasawa decomposition
$G^\bc =GAN$\cite{Zhel}. For $G^\bc=SL(N,\bc)$, the subgroup $AN$ is formed by the upper triangular complex matrices with positive real numbers on the diagonal and unit determinant.}  $D=G^\bc$, $\tilde G=AN$\ and $\E_\eta$ given by 
\be \E_\eta z =-z+\frac{1+\ri \eta }{2\ri\eta}\left((1+\ri \eta)z+(1-\ri\eta)z^*\right), \quad z\in\G^\bc\label{sbl}\ee
($z^*$ stands for the Hermitian conjugation). The
  ad-invariant non-degenerate symmetric bilinear form  $(.,.)_\D$  is given by the formula
\be (z_1,z_2)_{\G^\bc}:= -\ri(z_1,z_2)+\ri\overline{(z_1,z_2)}.\label{bgc}\ee
 
\noindent {\bf  5. The Poisson-Lie T-dual of the Yang-Baxter $\sigma$-model.} 
If, given an $\E$-model, there are two different subalgebras $\tilde\G_1$ and $\tilde\G_2$  having the properties described in {\bf 3.}  then the $\E$-model gives rise to two $\sigma$-models  living, respectively,  on different\footnote{However, if there exists an element $a\in D$ such that $Ad_a\tilde\G_1=\tilde\G_2$ then the target space geometries on $D/\tilde G_1$ and on $D/\tilde G_2$ are the same
in the sense of being related by a diffeomorphism   from $D/\tilde G_1$  onto  $D/\tilde G_2$.} targets $D/\tilde G_1$ and $D/\tilde G_2$. This phenomenon is called the Poisson-Lie T-duality and the models on $D/\tilde G_1$ and $D/\tilde G_2$ are referred to as being  Poisson-Lie T-dual to each other. Is there a Poisson-Lie T-dual to the Yang-Baxter $\sigma$-model \eqref{0}? Yes, there is, if we take $\tilde G_2=G$ instead of $\tilde G_1=AN$. The action of the dual $\sigma$-model  on the target $D/\tilde G_2$ 
in the form suitable for our exposition was worked out in \cite{K15} and it is given by the formula
\be\tilde S_\eta(p)= -\ri kS_{WZW}(p^2) -\ri k\int d\xi^+d\xi^- \left(\left(\frac{1+\ri \eta}{1-\ri \eta}- {\rm Ad}_{p^2}\right)^{-1}\d_+(p^2)p^{-2},p^{-2}\d_-(p^2)\right).\label{1}\ee
Here  the standard WZW action   $S_{WZW}(g)$  (based on the ordinary Killing-Cartan form $(.,.)$ on $\G^\bc$ and not on $(.,.)_{\D}$!) is given by
\be S_{WZW}(g):=  \frac{k}{2}\int  d\xi^+d\xi^-(g^{-1} \partial_+ g,g^{-1}\partial_-g)+ \frac{k}{12}\int d^{-1}(dgg^{-1},[dgg^{-1},dgg^{-1}])\ee
and $p(\xi^+,\xi^-)$ is a  field configuration taking values  in the space\footnote{For the group $G^\bc=SL(N,\bc)$, $P$ coincides with the space of positive definite Hermitian $N\times N$ matrices of unit determinant.}
 $P$ of positive definite Hermitian elements of $G^\bc$ which naturally parametrize the space of cosets $G^\bc/G$.  Note that the dual action $\tilde S_\eta(p)$ is real in spite of the occurence of the imaginary units
 in front of the integrals in the expression \eqref{1}.
 
 \vskip1pc
 
 \noindent {\bf  6. The  $\E$-model for the bi-Yang-Baxter $\sigma$-model.}
This paragraph {\bf 6.} interpolates between the review part of this letter presented so far and the original part to follow.  In fact, we expose here a result which is new, but could have 
been extracted without much difficulty from the contents of  Ref. \cite{K09}. Namely, we construct the $\E$-model corresponding to  the  two-parametric bi-Yang-Baxter $\sigma$-model  on the target of a simple compact group $G$  the second-order action of which reads
\be S_{\eta,\rho}(g)=  2k\eta\int  d\xi^+d\xi^-(g^{-1} \partial_+ g,(1-\eta R-\rho R_g)^{-1}g^{-1}\partial_-g).\label{0b}\ee
Here $R_g=$Ad$_{g^{-1}}R$Ad$_g$.  

To identify the $\E$-model from which \eqref{0b} originates we take, of course, the same double $D=G^\bc$  and the same subgroup $\tilde G_1=AN$ as in the case of the Yang-Baxter model \eqref{0}, however, the
crucial involution $\E_{\eta,\rho}$ must now be a one-parametric deformation of  the involution $\E_\eta$ from the paragraph {\bf 4}.  It turns out (at this is the first new result of this letter) that the correct choice is the following one
\be \E_{\eta,\rho} z=-z+\frac{1+\ri \eta +\rho R}{2\ri\eta}\left((1+\ri \eta-\rho R)z+(1-\ri\eta-\rho R)z^*\right), \quad z\in\G^\bc\label{sbr}\ee
where the operator $R$ is extended from $\G$ to $\G^\bc$ by complex linearity:
\be Rz:=\jp R(z-z^*)-\frac{\ri}{2}R(\ri z+\ri z^*).\ee
We now parametrize the coset $D/\tilde G_1=G^{\bc}/AN$ via the Iwasawa decomposition $G^\bc=GAN$ which means that the
configuration $f$ in Eq. \eqref{emo} is $G$-valued.  We set therefore $f=g$ and remark   that the term $S_{WZW,\D}(g)$ in \eqref{emo} vanishes because of the property of the Lie algebra $\tilde\G$ of $\tilde G=AN$ that $(u,u)_\D=0, \forall u\in\tilde\G$.
In order to see that the choice \eqref{sbr} gives the bi-Yang-Baxter model \eqref{1b}, it remains to identify the projection operator $P_{1,g}(\E_{\eta,\rho})$ on $\tilde\G_1$. 
For that, it helps to know that 
 every $\zeta\in\tilde\G$ can be uniquely written as
\be \zeta =(R-\ri)u\ee
for some $u\in\G$. With this insight, we find  that the following expression
\be P_{1,g}(\E_{\eta,\rho})z=\jp(R-\ri)(1+\rho R_g+\eta R)^{-1}\left((\ri+\ri \rho R_g+\eta)z+(\ri+\ri \rho R_g-\eta)z^*\right)\label{prr}\ee
verifies the conditions \eqref{nd} and,  inserting \eqref{prr} into \eqref{emo}, we recover the action \eqref{0b}.

 \vskip1pc
 
 \noindent {\bf  7. The Poisson-Lie T-dual of the bi-Yang-Baxter $\sigma$-model.} This is the central paragraph of this note since here we work out our  principal result which is the explicit form of the Poisson-Lie-T-dual of the bi-Yang-Baxter $\sigma$-model. Of course, the action of the dual $\sigma$-model is derived  from the  $\E$-model  based on $\D=\G^\bc$ and $\E_{\eta,\rho}$ via the basic formula \eqref{emo}, the thing which changes with respect to {\bf 6.}  is the choice of the Lie subgroup $\tilde G_2=G$.  Identifying the coset $D/G$   with the space $P$ of all positive definite Hermitian
elements of the group $G^\bc$ as in \cite{K15}, we set in \eqref{emo} $f=p\in P$ and find  the corresponding dual projection operator $P_{2,p}(\E_{\eta,\rho})$ on $\tilde \G_2=\G$ :
\be P_{2,p}(\E_{\eta,\rho})z=\jp(z-z^*)+ \frac{1}{2}(m_+(1+\rho R)-m_-\eta)(m_-(1+\rho R)+m_+\eta)^{-1}\ri(z+z^*),\label{pos}\ee
where  the operators $m_\pm:\G\to\G$ are defined by
\be m_+:=\jp({\rm Ad}_p+{\rm Ad}_{p^{-1}}), \quad m_-:= \frac{\ri}{2}({\rm Ad}_p-{\rm Ad}_{p^{-1}}).\ee Inserting \eqref{pos} in \eqref{emo} and realizing that $p$ is Hermitian (therefore it holds $(p^{-1}\partial_+p,p^{-1}\partial_-p)_\D=0$) we find after some work the action of the Poisson-Lie T-dual of the bi-Yang-Baxter $\sigma$-model:
$$\tilde S_{\eta,\rho}(p)=$$\be= -\ri S_{WZW}(p^2) -\ri k\int d\xi^+d\xi^- \left(\left(\frac{1+\ri \eta+\rho R}{1-\ri \eta+\rho R}- {\rm Ad}_{p^2}\right)^{-1}\d_+(p^2)p^{-2},p^{-2}\d_-(p^2)\right).\label{1b}\ee

 \noindent {\bf  8.  Generalized $\lambda$-model and the analytic continuation.}  Recently, Sfetsos, Siampos and Thomson introduced in \cite{SST15} an interesting two-parameter integrable deformation of the
 WZW model on the simple compact group $G$ which they called the  generalized  $\lambda$-model.  The action of this theory is given by the formula
\be S_{\alpha,\rho}(g) = S_{WZW}(g) +k\int d\xi^+d\xi^- \left(\left(\frac{1+\alpha+\rho R}{1-\alpha+\rho R}- {\rm Ad}_{g}\right)^{-1}\d_+gg^{-1},g^{-1}\d_-g\right),\label{8}\ee
 where $\alpha$,$\rho$ are real parameters related to the real  parameters  $\tilde t,\tilde\eta$ originally used  in \cite{SST15} by  the formulae
 \be \rho=-\frac{2k\tilde t\tilde\eta}{2k\tilde t+1}, \quad \alpha= \frac{1}{2k\tilde t+1}.\ee
 We remark also that the terminology $\lm$-model refers  to the notation used in \cite{SST15} where the operator $\frac{1+\alpha+\rho R}{1-\alpha+\rho R}$ was denoted as $\Lambda^{-1}$.
 
 It is now evident that the action 
  \eqref{8} can be transformed into that \eqref{1b} by performing three operations : 
  
  1)  replacing the $G$-valued configuration $g(\xi_+,\xi_-)$ by the $P$-valued $p^2(\xi_+,\xi_-)$, 
  
  2) replacing the real parameter $\alpha$ by the purely imaginary  one $i\eta$,
  
   3) multiplying the action \eqref{1b} by $-\ri$. 
   
The two last operations can be clearly interpreted as appropriate analytic continuations and the first one too, if we parametrize  $g$ and $p^2$ in the Cartan way:  $g$ as $g=k\delta k^{-1}$ and   $p^2$ as $p^2=kak^{-1}$ with $k\in G$,  $\delta$  is unitary diagonal  and $a$ is   real positive diagonal. Replacing  $\delta$  by $a$   can be now interpreted as a simple analytic continuation of the coordinates parametrizing the complex Cartan torus of $G^\bc$.
In the case of the target $SU(2)$, the  operations 1), 2) and 3)  coincide with those carried out in \cite{SST15} therefore 
our result  generalizes to any $G$ the $SU(2)$ result of Sfetsos, Siampos and Thompson stating that the  generalized $\lambda$-model  is related by  an appropriate analytic continuation to the Poisson-Lie T-dual of 
the bi-Yang-Baxter $\sigma$-model.  
 
\vskip1pc

  \noindent {\bf  9.  Poisson-Lie T-duals of the general KS-models.}
  Consider now the most general  $\E$-model  based on the  Drinfeld double $D=G^\bc$. It is defined by the choice of a linear operator $E:\G\to\G$, which can be written unambiguously as
  $E=S+A$, where $(Sx,y)_\G=(x,Sy)_\G$, $(Ax,y)_\G=-(x,Ay)_\G$, and we require also that  the
    symmetric part $S$ is invertible.
  We choose the corresponding  self adjoint involution $\E_{\eta,E}$   as
\be   \E_{\eta,E} z=-z+(E+i\eta)\frac{S^{-1}}{2\ri \eta}\left((E^\dagger+\ri\eta)z+(E^\dagger-\ri\eta)z^*\right),\quad z\in\G^\bc,\label{EE}\ee
  where $E^\dagger\equiv S-A$. Note that for $S$ equal to the  identity and $A=\rho R$ we recover the bi-Yang-Baxter involution $\E_{\eta,\rho}$.
  
  It is not difficult to work out the crucial projection operators. Setting $E_g:={\rm Ad}_{g^{-1}}E{\rm Ad}_g$, we find \be P_{1,g}(\E_{\eta,E})z=\frac{\ri}{2}(R-\ri)(E_g+\eta R)^{-1}\left((E_g-\ri\eta)z+(E_g+\ri\eta)z^*\right),  \ee
  \be P_{2,p}(\E_{\eta,E})z=\jp(z-z^*)+\jp(m_+E-m_-\eta)(m_-E+m_+\eta)^{-1}\ri(z+z^*)\ee
  which, plugged in the fundamental formula \eqref{emo}, yield respectively the generic KS-model 
  \be S_{\eta,E}(g)=  2k\eta\int  d\xi^+d\xi^-(g^{-1} \partial_+ g,(E^\dagger_g-\eta R)^{-1}g^{-1}\partial_-g).\label{0c}\ee
  and its Poisson-Lie T-dual
  $$\tilde S_{\eta,E}(p)=$$\be= -\ri S_{WZW}(p^2) -\ri k\int d\xi^+d\xi^- \left(\left(\frac{E+\ri \eta}{E-\ri \eta}- {\rm Ad}_{p^2}\right)^{-1}\d_+(p^2)p^{-2},p^{-2}\d_-(p^2)\right).\label{1e}\ee

 \noindent {\bf  10.  T-models and the analytic continuation.}   Replacing in \eqref{1e}  the $P$-valued $p^2(\xi_+,\xi_-)$ by the $G$-valued configuration $g(\xi_+,\xi_-)$, replacing the  imaginary parameter $\ri\eta$ by the real parameter $\alpha$       and multiplying the action \eqref{1e}        by  $\ri$, we obtain
\be S_{\alpha, E}(g)=  S_{WZW}(g) +k\int d\xi^+d\xi^- \left(\left(\frac{E+\alpha}{E-\alpha}- {\rm Ad}_{g}\right)^{-1}\d_+gg^{-1},g^{-1}\d_-g\right).\label{8x}\ee
  The action \eqref{1e}  is to be compared with  the general action of the T-model on the compact group target $G$:
\be S_{\Lambda}(g)= S_{WZW}(g) +k\int d\xi^+d\xi^- \left(\left(\Lambda^{-1}- {\rm Ad}_{g}\right)^{-1}\d_+gg^{-1},g^{-1}\d_-g\right),\label{8c}\ee
where $\Lambda:\G\to\G$ is an arbitrary invertible operator.
The obvious identification 
\be \Lambda^{-1}= \frac{E+\alpha}{E-\alpha}\ee
can be generically inverted 
\be E=-\alpha\frac{\Lambda+1}{\Lambda-1},\ee
 which confirms our claim that the analytical continuations of the Poisson-Lie T-duals of the KS-models are the T-models.

  \vskip1pc

  \noindent {\bf  11. Outlook.} In order to  relate the $\eta$ and the  $\lm$ deformations of the   $\sigma$-models  living on  the  {\it cosets} of $G$ via the Poisson-Lie T-duality and 
the analytic continuation, it looks promising to use the  framework  of
 the dressing cosets generalization of the $\E$-models  \cite{KS96b}.
  We plan to deal with this problem in a near future. Another interesting  question to study would  be the behavior  of the Poisson-Lie symmetries of the models \eqref{1e} under the analytic continuation
  yielding the T-models \eqref{8c}.
  The recent results of Ref. \cite{DLMV16b} could be of use in tackling this problem.
 
 \vskip2pc

  \noindent {\bf Acknowledgement}: I thank K. Siampos  for discussions and also for having urged me to work out the Poisson-Lie T-duals of the generic KS-model in the dual target space parametrization introduced
  in \cite{K15}.

\end{document}